\documentclass[aps,prl,twocolumn,,superscriptaddress,amssymb,amsmath]{revtex4}
\usepackage{graphicx}
\begin{document}

\title{State-independent proof of Kochen-Specker theorem with 13 rays}
\author{Sixia Yu}
\affiliation{Centre for quantum technologies, National University of Singapore, 3 Science Drive 2, Singapore 117543}
\affiliation{Hefei National Laboratory for Physical Sciences at
Microscale and Department of Modern Physics, University of Science
and Technology of China, Hefei, Anhui 230026, China}
\author{C.H. Oh}
\affiliation{Centre for quantum technologies, National University of Singapore, 3 Science Drive 2, Singapore 117543}

\begin{abstract}
Quantum contextuality, as proved by Kochen and Specker, and also by Bell, should manifest itself in any state in any system with more than two distinguishable states and recently has been experimentally verified on various physical systems. However for the simplest system capable of exhibiting contextuality, a qutrit, the quantum contextuality is verified only  state-dependently in experiment because too many (at least 31) observables are involved in all the known state-independent tests. Here we report an experimentally testable inequality involving only 13 observables that is satisfied by all non-contextual realistic models while being violated by all qutrit states. Thus our inequality will practically facilitate a state-independent test of the quantum contextuality for an indivisible quantum system. We provide also a record-breaking state-independent proof of the Kochen-Specker theorem with 13 directions determined by 26 points on the surface of a three by three magic cube.

\end{abstract}

\maketitle It is believed, almost religiously, that every effect has
its own cause and the same cause shall lead to the same effect. The
predictions of quantum mechanics (QM) are however probabilistic and
the effect that different outcomes appear in different runs of a
measurement seem to have no definite cause, at least unexplainable
using QM alone. Einstein, Podolsky, and Rosen \cite{Einstein}
initiated a longlasting quest for a quantum reality by questioning
the completeness of quantum mechanics.  Hidden variable (HV) models
are introduced in order to explain why a certain outcome appears in
each run of a measurement, attempting to make QM complete.  Years
later Kochen, Specker \cite{ks}, and Bell \cite{bell2} discovered
that quantum mechanics can be completed only by a hidden variable
model that is {\it contextual}: the outcome of a measurement depends on
which compatible observable might be measured alongside. Simply put,
Kochen-Specker (KS) theorem states that non-contextual HV models
cannot reproduce all the predictions of QM or quantum mechanics is
contextual.

In any non-contextual HV model all observables have definite values
determined only by some HVs $\lambda$ that are distributed according
to a given probability distribution $\varrho_\lambda$ with
normalization $\int d\lambda\varrho_\lambda=1$. Two observables are
compatible if they can be measured in a single experimental setup
and a maximal set of mutually compatible observables defines a
context. Non-contextuality is a typical classical property: the
value of an observable revealed by a measurement is predetermined by
HVs $\lambda$ only regardless of which compatible observable might
be measured alongside.  Local realism is a form of non-contextuality
enforced by the locality and thus Bell's inequalities \cite{bell}
are a special form of KS inequalities \cite{ksin1,ksin2,
5,cab1,pit}, experimentally testable inequalities that are satisfied
by all non-contextual HV models, some of which have been tested in
recent measurements \cite{zeil,ustc,yuji,yuji2,liu,siex, ex,cab2,3}.
In general KS inequalities reveal the nonclassical nature of single
systems demanding neither  space-like separation nor entanglement, i.e., independent of state.

\begin{figure}
\includegraphics{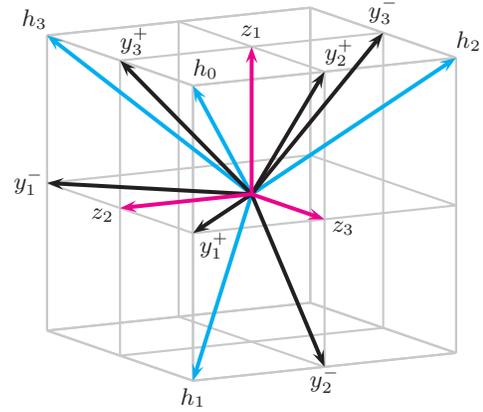}
\caption{(Color online) Illustration of 13 directions that are determined by 26 points on the surface of a $3\times 3$ magic cube. }
\end{figure}

However for the simplest system capable of exhibiting the quantum contextuality, a qutrit, only a state-dependent verification has been made in a recent experiment \cite{3}. This is because all the known state-independent KS inequalities for a qutrit \cite{pit} are based on the proofs of KS theorem and involve too many observables to be tested practically. For examples the original KS proof involves 117 directions \cite{ks} and Peres' \cite{peres} and Schuttle's \cite{schu} proofs involve 33 directions while the best KS proof known so far is due to Conway and Kochen \cite{31} in which 31 directions are still involoved. Also there are many state-dependent KS proofs among which a 5-direction proof \cite{5} has been used in the recent state-dependent experimental verification of quantum contextuality for qutrit. In this Letter we report a state-independent proof of KS theorem for qutrit with only 13 directions. Based on this proof we propose an experimental testable inequality that involves only 13 observables and two-observable correlations and is satisfied by all non-contextual HV models while being violate by all qutrit state. Thus our inequality will make a state-independent test of quantum contextuality for an indivisible system practical.

How can we exclude non-contextual HV models for QM or prove the quantum contextuality? Obviously the answer depends on what kinds of quantum mechanical predictions we want the HV model to reproduce. For example if we only want the predictions on non-sequential measurements, i.e., correlations not included,  to be reproduced, then a non-contextual HV model does exist according to Kochen and Specker \cite{ks}. Because of this toy model Kochen and Specker imposed a rather strong constraint on the HV models as a way out \cite{ks}: the algebraic structure of compatible observables must be preserved. Especially the value assigned to the product or the sum of two compatible observables must be equal to the product or the sum of the values assigned to these two compatible observables, which will be referred to as the product rule and the sum rule respectively. As we shall see later this constraint can be lifted if we consider sequential measurements.

As a result of the product rule the value assigned to the product of two orthogonal rays, normalized rank-1 projections, which are compatible, must be zero. As a result of the sum rule there is one and only one ray that is assigned to value 1 among all the rays in a complete orthonormal basis since the identity is always assigned to value 1. Thus in every non-contextual HV model preserving the partial algebraic structure of compatible observables there exists a {\it KS value assignment} to all rays in the corresponding Hilbert space satisfying:
\begin{enumerate}
\item The value $\{0,1\}$ assigned to a ray is independent of which bases it finds itself in;
\item One and only one ray is assigned to value 1 among all the rays in a complete orthonormal basis.
\end{enumerate}
The first condition reflects the non-contextuality and the second condition arises from the requirement that the algebraic structure of compatible observables be preserved. For a Hilbert space of a dimension greater than 2 there always exists a finite set of rays to which the  KS value assignment is impossible. For qutrits, a state-independent proof originally involves 117 rays \cite{ks} and the number is reduced to 33 by Peres \cite{peres} and Sch{\"u}tte as reported by K. Svozil in 1994 and pointed out by Bub \cite{schu}. The best KS proof known so far is given by Conway and Kochen \cite{31} with 31 rays. For 4-state systems the best state-independent proof is due to Cabello, Estebaranz, and Garc\'{\i}a-Alcaine  \cite{18} with 18 rays, the smallest state-independent KS proof known so far.

To warm up let us present a state-independent proof of KS theorem for qutrit using only 13 rays. In a given basis $\{|0\rangle,|1\rangle,|2\rangle\}$ we shall represent a qutrit ray $\hat r=|r\rangle\langle r|/\langle r|r\rangle$ by  a triple $r=(a,b,c)$ such that $|r\rangle=a|0\rangle+b|1\rangle+c|2\rangle$. Consider the following 13 rays
\begin{equation}\label{r}
\begin{array}{l@{\hskip 0.3cm}l@{\hskip 0.3cm}l}
y_1^-=(0,1,-1)&h_1=(-1,1,1) &z_1=(1,0,0) \\
y_2^-=(1,0,-1)&h_2=(1,-1,1) &z_2=(0,1,0)\\
y_3^-=(1,-1,0)& h_3=(1,1,-1)&z_3=(0,0,1) \\
y_1^+=(0,1,1)&h_0=(1,1,1)                  &\\
y_2^+=(1,0,1)&&\\
y_3^+=(1,1,0)&&\\
\end{array}
\end{equation}
that are determined by  26 points on the surface of a $3\times 3$ magic cube as illustrated in  Fig.1.  If we regard those 13 rays as 13 vertices and link two vertices if and only if the corresponding rays are orthogonal, then we obtain the orthogonality graph $\Delta_{13}$ as shown in Fig.2. Obviously a given set of rays determines uniquely the orthogonality graph and usually not vice versa. However those 13 rays are determined uniquely by the orthogonality relationships specified by the graph $\Delta_{13}$ up to a global unitary transformation.

\begin{figure}
\includegraphics{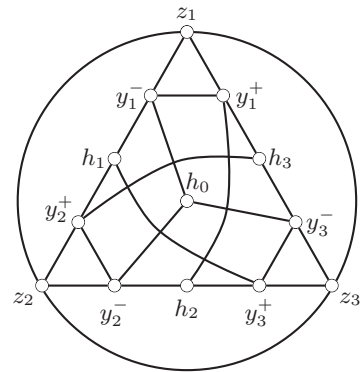}
\caption{ The orthogonality relationships among those 13 rays in
Eq.(\ref{r}) determine a graph $\Delta_{13}$ with 13 vertices
(hollow dots) representing those 13 rays and edges, straight or
curved, linking two rays that are orthogonal.}
\end{figure}

In fact without loss of generality we can choose $z_k$ as in Eq.(\ref{r})
since they form a basis. Because $\{z_k, y_k^\pm\}$ are mutually orthogonal for each $k=1,2,3$ there exist nonzero $t_1,t_2,t_3$ such that $y_1^+=(0,t_1,1)$ and $y_1^-=(0,-1,t_1^*)$, $y_2^+=(1,0,t_2)$ and $y_2^-=(t_2^*,0,-1)$,
$y_3^+=(t_3,1,0)$ and $y_3^-=(-1,t_3^*,0)$. As a result we have $h_1=(-t_2^*,t_1,1)$, $h_2=(1,-t_3^*,t_2)$, and $h_3=(t_3,1,-t_1^*)$. Since $h_k$ is orthogonal to $y_{k-1}^+$ for $k=1,2,3$ we have $t_1^*=t_2t_3$, $t_2^*=t_1t_3$, and $t_3^*=t_1t_2$ from which it follows that $|t_k|=1$ and $t_1t_2t_3=1$, i.e., $t_k=e^{i(\theta_{k+1}-\theta_{k+2})}$  for some real $\theta_k$. Finally we obtain $h_0=(e^{i\theta_1},e^{i\theta_2},e^{i\theta_3})$ which is orthogonal to $y_{1,2,3}^-$.
The diagonal unitary transformation taking $h_0$ to $(1,1,1)$ leaves $z_k$ unchanged so that the standard form of 13 rays in Eq.(\ref{r}) is obtained.

The KS value assignments to the 13-ray set are possible, i.e.,  no logical contradiction can be extracted by considering conditions 1 and 2 only. However in any possible KS value assignment there is at most one ray among $\{\hat h_\alpha|\alpha=0,1,2,3\}$ that can be assigned to value 1.  Suppose that this is not true, i.e., there are two or more $\hat h_\alpha$ that are assigned to value 1. Due to the symmetry of the graph $\Delta_{13}$ as shown in Fig.2, we need only to consider the following two cases:
\begin{itemize}
\item[i)] if $\hat h_0$ and $\hat h_1$ are assigned to value $1$ then $\hat y_2^\pm$ and $\hat y_3^\pm$ must be assigned to 0 so that both $\hat z_2$ and $\hat z_3$ must be assigned to value 1 which is impossible;
\item[ii)] If $\hat h_1$ and $\hat h_2$ are assigned to 1 then  $\hat y_1^\pm$ and $\hat y_2^\pm$ must be zero so that both $\hat z_1$ and $\hat z_2$ must be assigned to value 1, also a contradiction.
\end{itemize}
In the reasonings above we have taken into account of condition 2 which demands that linked rays not be assigned simultaneously to value 1 and in a triangle one and only one ray be assigned to value 1. A set of eight rays in each case considered above, e.g., $\hat h_{0,1}$, $z_{2,3}$, and $y_{2,3}^\pm$, constitutes in fact a 3-box paradox \cite{3box} which appears also in Cliffton's  state-dependent proof of KS theorem \cite{clif} and a recent proposal to close the compatibility loopholes \cite{cb}.

Denoting by $h^\lambda_\alpha\in\{0,1\}$ the value assigned to $\hat h_\alpha$ for given $\lambda$, we then have $\sum_\alpha h_\alpha^\lambda\le 1$ from the above arguments. As a result the following inequality
\begin{equation}\label{4}
\sum_{\alpha=0}^3 \langle \hat h_\alpha\rangle_c:=\sum_{\alpha=0}^3 \int d\lambda \varrho_\lambda h_\alpha^\lambda\le 1
\end{equation}
must be satisfied by all non-contextual HV models that admit a KS value assignment. However the quantum mechanics predicts $\sum_{\alpha=0}^3\langle \hat h_\alpha\rangle_q=4/3
$ due to the identity $\sum_{\alpha=0}^3 \hat h_\alpha=\frac 43 I$ with $I$ being the identity operator of qutrit (see Appendix). This proves the original KS theorem: the non-contextual HV model satisfying conditions 1 and 2 cannot reproduce all the predictions of QM on non-sequential measurements.

Usually the KS theorem is proved by finding a set of rays to which the KS value assignment does not exist so that we need not to check other predictions of QM. Our proof here is a set of 13 rays, to which all possible KS value assignments, which do exist, fail to reproduce a certain prediction of QM. Notably the inequality Eq.(\ref{4}), involving only 4 observables explicitly, provides a state-independent test for the non-contextual HV models that preserve the algebraic structure of compatible observables, in the spirit of Kochen and Specker. In an experimental test of the inequality Eq.(\ref{4}) each average must be measured on different subensembles prepared in the same state, similar to a standard test of a Bell inequality as pointed out by Cabello \cite{cab1}.

It turns out that the requirement of preserving the algebraic structure, i.e., the product and sum rules, of compatible observables is too strong and unnecessary. Peres already noticed that the sum rule can be abandoned and kept the product rule \cite{peres4} as in KS proofs via the Mermin-Peres square \cite{peres,m2} for 2 qubits and Mermin's pentagram \cite{m2} for 3 qubits. However the product rule is not necessary either. Instead we need only to require that the quantum correlations of compatible observables, the  quantum mechanical predictions on sequential measurements, be reproduced. This imposes no additional constraints on the non-contextual HV models since they must reproduce all the predictions of QM in the first place. As noticed earlier, the toy model by Kochen and Specker did not take into account of the quantum correlations of compatible observables.

Indeed every KS proof mentioned previously can be turned into a state-independent KS inequality \cite{pit} that is obeyed by all non-contextual HV models, no matter whether the algebraic structure of compatible observables is preserved or not. All known state-independent KS inequality involve the correlations of at least three compatible observables, i.e., predictions on three or more sequential measurements must be reproduced by the HV models. In the case of qutrit the resulting KS inequalities involve too many observables, e.g., at least 31, to be tested experimentally.  Recently Klyachko, Can,  Binicioglu, and Shumovskya \cite{5} propose a simple KS inequality, called pentagram inequality since it is based on the graph of a pentagon, to test non-contextual HVs with only 5 dichotomic observables and correlations of two compatible observables. However this pentagram inequality, which is based on the assumption of non-contextuality only and whose  violation has been verified in a recent experiment \cite{3},  has the drawback of being state-dependent.

Our state-independent inequality is based on the orthogonality graph $\Delta_{13}$ as shown in Fig.2. We denote  by $V=\{y_k^\sigma,h_\alpha,z_k|k=1,2,3;\sigma=\pm;\alpha=0,1,2,3\}$ its vertex set and by  $\Gamma$ Its adjacency matrix which is a  $13\times 13$ symmetric matrix with vanishing diagonal elements and $\Gamma_{uv}=1$ if two vertices $u,v\in V$  are neighbors and $\Gamma_{uv}=0$ otherwise. For arbitrary 13 variables $a_v=\pm1$ with $v\in V$ it holds
\begin{equation}\label{1}
\sum_{v\in V}a_v-\frac14\sum_{u,v\in V}\Gamma_{uv}a_u a_v\le 8,
\end{equation}
which can be easily verified  with the help of a laptop by exhausting all $2^{13}$ possibilities and an analytic proof is provided in Appendix. 
Let $\{A_v| v\in V\}$ be a set of 13 dichotomic observables taking values $a_v^\lambda=\pm1$ for given $\lambda$. Then from inequality Eq.(\ref{1}) we obtain our magic-cube inequality
\begin{equation}\label{ineq}
\sum_{v\in V}\langle A_v\rangle_c-\frac14\sum_{u,v\in V}\Gamma_{uv}\langle A_u\cdot  A_v\rangle_c
\le 8,
\end{equation}
where we have denoted $\langle A_v\rangle_c:=\int d\lambda \varrho_\lambda a^\lambda_v$ and $\langle A_u\cdot A_v\rangle_c:=\int d\lambda \varrho_\lambda a_u^\lambda a_v^\lambda
$. Though the correlation of two observables is always well defined  in a non-contextual HV model regardless of whether they are compatible or not, in order to compare with the predictions of QM those observables labeled with linked vertices in the orthogonality graph $\Delta_{13}$ should be compatible, i.e, they can be measured simultaneously or be commuting, so that their correlation is also well defined in QM.

Fortunately if we define 13 observables $\hat A_v=I-2\hat r_v$ from 13 rays $\hat r_v\in\{\hat y_k^\sigma,\hat h_\alpha,\hat z_k\}$, then $\hat A_u$ and $\hat A_v$ are commuting. i.e., compatible, whenever two vertices $u,v\in  V$ are linked, i.e., $\Gamma_{uv}=1$. Therefore all the expectation values appear on the left-hand-side of Eq.(\ref{ineq}) are also well defined in QM and therefore should be reproduced. Due to the linearity of QM the left-hand-side of Eq.(\ref{ineq}) must reproduce the expectation value of
\begin{eqnarray}
\hat  L&=&\sum_{v\in V}\hat A_v-\frac14\sum_{u,v\in V}\Gamma_{uv}\hat A_u \hat A_v.
\end{eqnarray}
Simple calculations yield $\hat L=\frac{25}3 I$ and therefore $\langle \hat L\rangle_{q}=25/3>8$ for any qutrit state, meaning that the magic-cube inequality Eq.(\ref{ineq}) is violated by all qutrit states. Thus we have proved in a state-independent fashion that every non-contextual HV model, no matter whether the algebraic structure of compatible observables is preserved or not, cannot reproduce all the predictions of QM, especially those quantum correlations of two compatible observables, by using only 13 observables.

To summarize, our magic-cube inequality Eq.(\ref{ineq}) provides the simplest state-independent proof of quantum contextuality: Firstly it is a test for the contextuality for an indivisible system capable of exhibiting the quantum contextuality, i.e., smallest system; Secondly it involves a smallest number of observables so far and we conjecture that KS theorem does not have a state-independent proof with less than 13 rays; Thirdly only correlations of at most two, the smallest number,  of compatible observables are involved so that only two observables have to be measured sequentially.  In comparison correlations of at least three compatible observables are involved in all the state-independent tests of contextuality known so far. If we impose a minimal requirement on the non-contextual HV models, i.e,  only the correlations of at most two compatible observables  are required to be reproduced, then our magic-cube inequality will be the first state-independent proof of KS theorem in this case. We believe that there are also proofs of KS theorem for higher dimensional systems that involve only two-observable correlations. Finally  we have proved the KS theorem by finding a 13-ray set, which is the smallest state-independent proof so far, for which all possible KS value assignments, which do exist, fail to reproduce a certain prediction of QM. Since only a relatively small number of observables and correlations of only two compatible observables are involved, an experimental test of our inequality is well within the reach of the current technologies. We believe that the nitrogen-vacancy center in diamond \cite{nv} should be a promising choice to test our inequality.

This work is supported by National
Research Foundation and Ministry of Education, Singapore (Grant No.
WBS: R-710-000-008-271) and NSF of China (Grant No. 11075227).


\newpage

\begin{widetext}
\section*{Appendix}
{\it Explicit rays for Eq.(\ref{r})--- } By denoting $\bar 1=-1$ we have
\begin{equation*}
\hat y_1^\pm=\frac12\left(\begin{array}{ccc}
0&0&0\\0&1&\pm1\\0&\pm1&1
\end{array}\right),\
\hat y_2^\pm=\frac12\left(\begin{array}{ccc}
1&0&\pm1\\0&0&0\\\pm1&0&1
\end{array}\right),\
\hat y_3^\pm=\frac12\left(\begin{array}{ccc}
1&\pm1&0\\\pm1&1&0\\0&0&0
\end{array}\right),
\end{equation*}
\begin{equation*}
\hat h_0=\frac13\left(\begin{array}{ccc}
1&1&1\\1&1&1\\1&1&1
\end{array}\right),\
\hat h_1=\frac13\left(\begin{array}{ccc}
1&\bar1&\bar1\\\bar1&1&1\\\bar1&1&1
\end{array}\right),\
\hat h_2=\frac13\left(\begin{array}{ccc}
1&\bar1&1\\\bar1&1&\bar1\\1&\bar1&1
\end{array}\right),\
\hat h_3=\frac13\left(\begin{array}{ccc}
1&1&\bar1\\1&1&\bar1\\\bar1&\bar1&1
\end{array}\right),
\end{equation*}
\begin{equation*}
\hat z_1=\left(\begin{array}{ccc}
1&0&0\\0&0&0\\0&0&0
\end{array}\right),\
\hat z_2=\left(\begin{array}{ccc}
0&0&0\\0&1&0\\0&0&0
\end{array}\right),\
\hat z_3=\left(\begin{array}{ccc}
0&0&0\\0&0&0\\0&0&1
\end{array}\right),\
I=\left(\begin{array}{ccc}
1&0&0\\0&1&0\\0&0&1
\end{array}\right).
\end{equation*}
Thus we have identities
\begin{eqnarray}
\hat y:=\sum_{k=1}^3\sum_{\sigma=\pm}\hat y_k^\sigma=2I,\quad
\hat h:=\sum_{\alpha=0}^3\hat h_\alpha=\frac 43I,\quad
\hat z:=\sum_{k=1}^3\hat z_k=I.
\end{eqnarray}
Let $\hat A_v=I-2\hat r_v$ with $\hat r_v\in \{\hat y_k^\sigma,\hat h_\alpha,\hat z_k\}$. Since
$24=\sum_{u,v\in V}\Gamma_{uv}/2$ is the number of the edges in $\Delta_{13}$, $\sum_{u\in V}\Gamma_{uv}$ is the degree of the vertex $v$, and $\hat r_v \hat r_u=0$ if $\Gamma_{uv}=1$, it holds
\begin{eqnarray}
\hat  L&=&\sum_{v\in V}\hat A_v-\frac14\sum_{u,v\in V}\Gamma_{uv}\hat A_u \hat A_v\cr
&=&13I-2(\hat y+\hat h+\hat z)-\frac I4\sum_{uv}\Gamma_{uv}+\sum_{uv}\Gamma_{uv} \hat r_v-\sum_{u,v\in
V}\Gamma_{uv}\hat r_u \hat r_v\cr
&=&13I-2(\hat y+\hat h+\hat z)-12I+4(\hat z+\hat y)+3\hat h\cr
&=&I+2(\hat z+\hat y)+\hat h=\frac {25}3 I.
\end{eqnarray}

{\it Proof of Eq.(4)--- } There are 9 vertices $\{z_k,y_k^\sigma\}$ of degree 4 and 4 vertices $\{h_\mu\}$ of degree 3 in $\Delta_{13}$, where the degree of a vertex denotes the number its neighbors.
Let $t$ be the total number of $a_v$ that take value $-1$,  $f$ be the number of $a_v$ that take value $-1$ with $v$ being of degree 4, and $l$ be the number of pairs of vertices $u,v$ such that $a_u=a_v=-1$ and $\Gamma_{uv}=1$. We then have $L=1+t+f-2l$. Since the quadratic term in $L$ is unchanged under replacements $a_v\mapsto -a_v$, we need only to consider $t\le 6$.  If $t\le 3$, since $\l\ge 0$ and $f\le t$, then we have $ L\le 7$. If $t=4$, then from $f-2l\le 3$ since it is impossible to have $f=4$ and $l=0$, i.e., among any 4 vertices of degree 4 there is at least a connected pair, it follows $ L\le 8$, which is attained when $f=3$ and $l=0$, e.g., $z_1, y_2^-,y_3^+$ and $h_3$. In the case of $t=5$, if $f=5$ then $l\ge 2$ so that $ L\le 7$. If $f\le 4$ then $l\ge 1$ so that $ L\le 8$, which is attained by, e.g., $z_1, y_2^-,y_3^+,y_1^-$ and $h_3$. In the case of $t=6$, if $f=6$ then $l\ge 3$ so that $ L\le 7$; If $2\le f\le 5$ then we have again $ L\le 8$ because $l\ge 2$; If $f=1$ then $L\le 7$ since $l\ge 0$.

\end{widetext}


\begin{thebibliography}{99}
\bibitem{Einstein} A. Einstein, B. Podolsky, and N. Rosen, 
Phys. Rev. {\bf 47}, 777–780 (1935).
\bibitem{ks} S. Kochen and E.P. Specker, J. Math. Mech. {\bf 17}, 59 (1967).
\bibitem{bell2} J.S. Bell, Rev. Mod. Phys. {\bf 38}, 447 (1966).
\bibitem{bell} J.S. Bell, Physics (Long Island City, N.Y.) {\bf 1}, 195 (1964).
\bibitem{ksin1} C. Simon, \v{C} Brukner, and A. Zeilinger, Phys. Rev. Lett. {\bf 86}, 4427 (2001).
\bibitem{ksin2} J.{\AA}. Larsson, Europhys. Lett. {\bf 58}, 799 (2002). \bibitem{5} A.A. Klyachko, M.A. Can, S. Binicioglu, and A.S. Shumovsky, Phys. Rev. Lett. {\bf 101}, 020403 (2008).
\bibitem{cab1} A. Cabello, 
Phys. Rev. Lett. {\bf 101}, 210401 (2008).
\bibitem{pit} P. Badzi\c{a}g, I. Bengtsson, A. Cabello, and I. Pitowsky, Phys. Rev. Lett. {\bf 103}, 050401 (2009).
\bibitem{zeil} M. Michler, H. Weinfurter, and M. Zukowski, Phys. Rev. Lett. {\bf 84}, 5457 (2000).
\bibitem{ustc} Y.F. Huang, C.F. Li, Y.S. Zhang, J.W. Pan, and G.C. Guo, Phys. Rev. Lett. {\bf 90}, 250401 (2003).
\bibitem{yuji}Y. Hasegawa, R. Loidl, G. Badurek, M. Baron, and H. Rauch, Phys. Rev. Lett. {\bf 97}, 230401 (2006).
\bibitem{yuji2} H. Bartosik, {\it etal.}, 
Phys. Rev. Lett. {\bf 103}, 040403 (2009).
\bibitem{liu} B.H. Liu, {\it et al.}, 
 Phys. Rev. A {\bf 80}, 044101 (2009).
\bibitem{siex} G. Kirchmair, 
{\it et al.}, Nature {\bf 460}, 494 (2009).
\bibitem{cab2} E. Amselem, M. R{\aa}dmark, M. Bourennane, and A. Cabello, 
Phys. Rev. Lett. {\bf 103}, 160405 (2009).
\bibitem{ex} O. Moussa, C.A. Ryan, D.G. Cory, and R. Laflamme, Phys. Rev. Lett. {\bf 104}, 160501 (2010).
\bibitem{3}R. Lapkiewicz, {\it etal.}, Nature {\bf 474}, 490 (2011).



\bibitem{peres} A. Peres, J. Phys. A {\bf 24}, L175 (1991).
\bibitem{schu} J. Bub, Found. Phys. {\bf 26}, 787 (1996).
\bibitem{31}J.H. Conway and S. Kochen around 1990, first reported in A. Peres, {\it Quantum theory: Concepts and Methods} (Kluwer, Dordrecht, 1993), p. 114. See also J.H. Conway and S. Kochen, in {\it Quantum [Un]speakables: From Bell to quantum information}, edited by R.A. Bertlmann and A. Zeilinger (Springer-Verlag, Berlin, 2002), p. 257.
\bibitem{18}A. Cabello, J.M. Estebaranz, G. Garc\'{\i}a-Alcaine, Phys. Lett. A {\bf 212}, 183 (1996).
\bibitem{3box} D.Z. Albert, Y. Aharonov, and S. D'Amato, Phys. Rev. Lett. {\bf 54}, 5 (1985).
\bibitem{clif} R. Clifton, 
Am. J. Phys. {\bf 61}, 443 (1993). 
\bibitem{cb} A. Cabello and M.T. Cunha, 
Phys. Rev. Lett. {\bf 106}, 190401 (2011).
\bibitem{m2} N.D. Mermin, Phys. Rev. Lett. {\bf 65}, 3373 (1990).
\bibitem{peres4} A. Peres, Phys. Lett. A {\bf 151}, 107 (1990).
\bibitem{nv}J.R. Maze, {\it etal.}, New J. of Phys. {\bf 13}, 025025 (2011).

\end{thebibliography}
\end{document}